# Qualitative observation of reversible phase change in astrochemical ethanethiol ices using infrared spectroscopy

S Pavithraa[a], R Rajan[b], P Gorai[c], J -I Lo[d], A Das[c], B N Raja Sekhar[e], T Pradeep[b], B -M Cheng[c], N J Mason[f] and B Sivaraman[a, *]

[a]Atomic Molecular and Optical Physics Division, Physical Research Laboratory, Ahmedabad, India.

[b]Department of Chemistry, Indian Institute of Technology-Madras, Chennai, India.

[c]IndianCenter for Space Physics, Kolkatta, India.

[d]National Synchrotron Radiation Research Center, Hsinchu, Taiwan.

[e]B-1, Indus-1, Atomic Molecular Physics Division, BARC at RRCAT, Indore, India.

[f]Department of Physical Sciences, The Open University, Milton Keynes, UK.

*Email: bhala@prl.res.in

**Abstract**

Here we report the first evidence for a reversible phase change in an ethanethiol ice prepared under astrochemical conditions. InfraRed (IR) spectroscopy was used to monitor the morphology of the ice using the S-H stretching vibration, a characteristic vibration of thiol molecules. The deposited sample was able to switch between amorphous and crystalline phases repeatedly under temperature cycles between 10 K and 130 K with subsequent loss of molecules in every phase change. Such an effect is dependent upon the original thickness of the ice. Further work on quantitative analysis is to be carried out in due course whereas here we are reporting the first results obtained.

**Introduction**

Since the discovery of methanol and hydrogen sulphide in the InterStellar Medium (ISM) in the early 1970's it was expected that that thiols should be a component of interstellar heterogeneous chemistry at low temperatures and that such thiols may, under energetic processing, lead to the synthesis of many more complex molecules[1-2]. Subsequently several sulphur bearing molecules have been discovered including the very first identification of methanethiol[3] and thiocyanic acid (HSCN); [4] in the InterStellar Medium (ISM). Ethanethiol was identified in the ISM in 2014 by Kolesniková, Tercero et al[5] and the presence of larger thiol groups, such as propanethiol ($CH_3CH_2CH_2SH$) is expected from astrochemical models, although in concentrations that are insufficient for detection using current ground based facilities [6].

The morphology of ices formed in astrochemical environments may be different from those formed under terrestrial conditions due to the low pressures and the long times needed for accretion of material in the ISM [7]. Molecular oxygen exists in many crystalline phases, α, β, and γ [8], whereas in the case of simulating oxygen ices at astrochemical conditions only two phases α, β [9] of oxygen ices exist. The morphology is also expected to change as the ice is subjected to thermal and energetic particle processing[5].

The morphology of molecular ices is an important parameter in the physical and chemical features of the ice, for example it has a strong influence on the porosity of the ice. Amorphous ice is, in general, more porous than crystalline ices. In a mixture of molecular ices, for example, in the case of acetonitrile-water mixture, when acetonitrile turns crystalline well before water crystallization temperature, the water molecules are segregated from the mixture[10]. Therefore amorphous ice, due to its higher porosity, relative to crystalline ice, acts as a trap for other molecules which in turn strongly influences the chemical reactivity of the ice at a given temperature.

The morphology of an ice may be explored by studying its spectroscopy and whilst VUV spectra[11] may show distinct changes when transforming from one phase to another such phase changes are more commonly seen by monitoring the InfraRed (IR) spectra of these ices which reflect the changes in the vibrational frequencies of the molecules in different phases of the ice[12].

To date phase change in such ices have been found to be irreversible with amorphous ice prepared at very low temperatures, such as 10 K, and then warmed to higher temperatures turning crystalline and remaining crystalline until sublimation. For example in a very recent work, methanethiol IR spectra revealed phase changes to have occurred while warming the ice from 10 K to higher temperatures[13]. In this paper we report the first experiments exploring ethanethiol molecular ices under astrochemical conditions and study changes in the ice morphology throughout thermal processing cycles which repeatedly heat and cool the ice providing the first evidence for reversible phase changes by thermal cycling.

**Experimental methodology**

IR spectra of condensed ethanethiol were measured using a Fourier Transform IR (FTIR) spectroscopy technique. Liquid ethanethiol samples (sigma Aldrich, purity 99.99%) were placed in a glass bulb and connected to the inlet gas line within which it was further purified by three freeze pump thaw cycles. Solid ethanethiol was vapor deposited directly on a pre-cooled potassium bromide (KBr) window mounted on a closed cycle helium cryostat (Janis RDK-415) which can be cooled to 4 K inside an ultra-high vacuum chamber. The temperature of the window was measured using a silicon diode temperature sensor. Spectra were recorded at various temperatures (4 K to 130 K) in the mid IR range (500-5000 $cm^{-1}$) with 2 $cm^{-1}$ resolution using a Bomem DA8 FTIR equipped with a KBr beam splitter and a MCT detector (cooled by liquid nitrogen); in which, the interferometric spectrometer is coupled to the photochemical end-station[14] attached to the beamline BL21A2 at National Synchrotron Radiation Research Center (NSRRC). The temperature of the ice could be changed by heating the KBr window using a resistive heater, controlled by a Lakeshore temperature controller (model 340). In these experiments the sample was heated to 100 K at 5 K $min^{-1}$ ramping rate and spectra were collected every 10 K intervals. Above 100 K the ice was annealed at a lower rate at 3 K $min^{-1}$ and spectra recorded at 110 K, 120 K, 125 K and 130 K. The experiment was then repeated with different ice thicknesses. Since the A (cm $molecule^{-1}$) value and densities of these ices are not known absorbance in the IR spectra was used for monitoring the ice thickness which is of the order of a few microns.

**Results and Discussion**

A thin film of ethanethiol ice was formed at 4 K at a pressure of 1 x $10^{-8}$ mbar for 15 secs and heated in steps of 10 K from 10 K. The spectra obtained at different temperatures are shown in Figure 1a. Peak assignments are given in Table 1. The peak at 2975 $cm^{-1}$ and 2959 $cm^{-1}$ are from $CH_2$ asymmetric and $CH_3$ asymmetric stretching vibrations. The S-H vibrations appear at 2528 $cm^{-1}$. The peaks at 1448 $cm^{-1}$ and 1430 $cm^{-1}$ are from $CH_3$ bending and $CH_2$ scissoring vibrations. The features at 1250 $cm^{-1}$ and 1094 $cm^{-1}$, 1050 $cm^{-1}$ correspond to $CH_2$ twisting, $CH_3$ bending and $CH_2$ rocking modes while C-C stretching is seen at 969 $cm^{-1}$ and C-S stretching at 655 $cm^{-1}$. Only small changes in the spectra were observed with the S-H stretching vibration shifting slightly from 2528 $cm^-$ at 4 K to 2536 cm$^{-1}$ at 130 K with a decrease in intensity (Figure 1b). Above 130 K ethanethiol peaks were not observed suggesting the ice has sublimed.

By depositing at a pressure averaged to about 5.5 x 10$^{-8}$ mbar for 60 secs to form a thick layer of ethanethiol ice at 4 K and recording the IR spectra similar band positions were observed as in the thin ethanethiol ice. Once again warming the ice to higher temperatures showed no significant change up to 100 K. However, upon increasing the temperature to 110 K, the S-H stretching vibration at 2528 cm$^{-1}$ was observed to split to two peaks as 2526 cm$^{-1}$ and 2541 cm$^{-1}$ (Figure 2a). The 1430 cm$^{-1}$ band corresponding to CH$_2$ scissoring vibration was observed to split into two peaks centred at 1425 cm$^{-1}$ and 1428 cm$^{-1}$. The band centred at 1094 cm$^{-1}$ representing the CH$_2$ twisting and rocking modes split to two peaks one at 1099 cm$^{-1}$ and the other at 1092 cm$^{-1}$. The band corresponding to SH out of plane bending at 864 cm$^{-1}$ was split to two bands one at 865 cm$^{-1}$ and the other at 874 cm$^{-1}$. Bands corresponding to CH$_2$ rocking and C-S stretching observed at 733 cm$^{-1}$ and 655 cm$^{-1}$ were observed to have split to 741 cm$^{-1}$, 730 cm$^{-1}$ and 656 cm$^{-1}$, 654 cm$^{-1}$, respectively (Table 1; Figure 2a). From these spectra we conclude that a phase transition from amorphous to crystalline phase had occurred in the ethanethiol ice. By re-cooling the sample from 120 K to 4 K the crystallinity was found to be retained (Figure 2b).

Upon warming the ice from 120 K to 125 K and with subsequent recording of IR spectra we found that the spectrum recorded was different to that spectrum at 120 K. The S-H stretching vibration was again found to centre at 2538 cm$^{-1}$ and no splitting was observed. This is also the case for all the other bands corresponding to the CH$_3$ bending, C-C stretching, CH$_2$ rocking and C-S stretching indeed the spectrum at 125 K resembled the spectra recorded from 4 K – 100 K (Figure 3a). This remarkable transformation from crystalline phase to amorphous phase at higher temperatures is observed for the first time in any astrochemical ices experiment.

By cooling the sample from 125 K to 4 K and then warming it again with subsequent recording of IR spectra we could clearly see from the S-H, CH$_2$, C-C and C-S characteristic vibrations that the sample had once again turned crystalline in the temperature window 110 K – 120 K (Figure 3a). The small peak shift observed in the S-H vibration by recooling the sample to 4 K could be due to the conformational changes in the ethanethiol homo-dimers between the sample deposited at 4 K and the sample recooled from higher temperature. Upon warming to 125 K the sample was once again found to turn from crystalline to amorphous phase. Now cooling the sample to lower temperatures and then back to 120 K we found that the sample remained amorphous at all temperatures (Figure 3b). This is due to the loss of molecules during every phase change reducing the ice thickness which now shows identical results to those shown in Figure 1. Loss of molecules from the ice phase during every phase change could be clearly observed by the observation of a rise in chamber pressure.

In order to confirm these results even thicker ethanethiol ice was deposited at 4 K and at a pressure of 5 x 10$^{-7}$ mbar for 30 sec. Upon warming this ice to 110 K we once again observe the band splitting indicating the phase change from amorphous to crystalline. A spectrum recorded at 125 K was again found to resemble the spectrum recorded at lower temperatures indicating the phase change from crystalline to amorphous. As we re-cooled the sample from 126 K to lower temperatures a spectrum was recorded at 118 K. This spectrum was surprising as we found band splitting to have happened indicating the phase change to have occurred from amorphous to crystalline upon cooling the sample (Figure 4). This is the first time evidence for crystallize upon cooling from higher to low temperatures is observed in astrochemical ices [15]. Further repeating the temperature cycle in the 100 K to 130 K region we found that the crystallization from re-cooling from higher to lower temperature was repeated.

**Conclusion**

We have performed, for the first time, IR spectroscopic studies of ethanethiol ices relevant to icy mantles of interstellar cold dust. Ethanethiol ices were formed at 4 K on a cold substrate and temperature dependent IR spectra were recorded. From the spectral changes observed we conclude that phase transition had occurred in the ethanethiol ice. Thin samples of ethanethiol ice were found to desorb without undergoing any phase transition, remaining amorphous, from 4 K until 130 K, before sublimation. However, by increasing the ethanethiol ice thickness a phase change from amorphous to crystalline ice at 110 K was observed. By further warming the crystalline sample to 125 K we observed, for the first time, that the morphology changed from crystalline to amorphous upon heating the sample. When re-cooling the amorphous sample from 125 K to lower temperatures and then heating again to 110 K the ice turned amorphous to crystalline and then from crystalline to amorphous at 125 K. This was repeated until the sample became thin enough from sublimation of molecules during every phase change to revert to the condition in which no changes in morphology is seen during the temperature cycle.

For an even thicker sample the amorphous to crystalline phase change was observed at 110 K and then from crystalline to amorphous phase at 125 K, however, while re-cooling this thicker sample from 125 K to 118 K another phase change from amorphous to crystalline was observed. This is the first time an astrochemical ice is known to change phase from amorphous to crystalline by cooling the sample. The temperature cycle of this sample was found to repeat the phase changes between amorphous and crystalline until the thickness is reduced further due to sublimation during every phase change.

These results have implications in both the ethanethiol chemistry in the cold dust grains of the interstellar medium where thiols are known to take part in reactions that could lead to the complex molecules and in the use of morphology as a monitor of thermal history of ices in the ISM (and other astronomical environments).

**Acknowledgement**

BS, SP and NJM would like to acknowledge the support from Sir John Mason Academic Trust and a beamtime grant from NSRRC, Taiwan. BS and SP would like to thank the DST INSPIRE grant (# IFA 11CH-11). The authors would like to acknowledge the support from Mr. SaravanaPrashanthMuraliBabu, Graduate Student, National Chiao Tung University, Hsinchu, Taiwan. RRJM and TP acknowledge the support from IITM. NJM and BS acknowledge the support from The Open University, UK. PG want to acknowledge the DST (Grant No. SB/S2/HEP-021/2013) project for his financial support and AD want to acknowledge ISRO respond (Grant No. ISRO/RES/2/402/16-17) project.

**Figure captions**

**Figure 1: (a)** An IR spectrum recorded for a thin layer of pure ethanethiol molecules deposited at 4 K and subsequently warmed to higher temperatures. Ethanethiol remains in an amorphous phase and sublimes at 130 K without undergoing a phase change. **(b)** The observed shift in the peak position and decrease in intensity of the S-H stretching region in thin layer deposition of ethanethiol ices as the molecules are warmed from 4 K to 135 K.

**Figure 2:** Infrared spectra of; **(a)** a pure ethanethiol ice formed at 4 K. (Top) No phase change in thin ethanethiol ice is observed upon warming to higher temperature. (Bottom) a thick ethanethiol ice undergoing phase change at 110 K and reversible phase change from crystalline to amorphous upon further warming to 130 K. **(b)** a thick ethanethiol ice formed at 4 K, warmed to 120 K and cooled back to 4 K to confirm the crystalline phase and warmed again to 120 K. The figure is blown up and presented in two sets of wavenumbers i.e. CH str and SH str region (top), fingerprint region (bottom), so that the spectral features are clearly visible. Note: Black - amorphous, Blue - crystalline phase.

**Figure 3:** Infrared spectra of *(a)* a thick ethanethiol ice formed at 4 K (amorphous), warmed to 120 K (crystalline), further warming to 125 K (amorphous), cooled back to 4 K, again warmed to 100 K and 118 K (crystalline). Note: Black - amorphous, Blue - crystalline phase.(b) An experiment to measure the thickness dependence of the phase change. CYCLE 1: Thick ethanethiol ice deposited at 4 K and subsequently warmed to 110 K, 120 K (crystalline) and 125 K (amorphous). CYCLE 2: The ice at 125 K in CYCLE 1 was cooled back to 4 K and again warmed to 120 K (crystalline) and 125 K (amorphous). CYCLE 3: The ice at 125 K in CYCLE 2 was cooled to 100 K and subsequently warmed to 120 K (amorphous), 125 K (amorphous). Note: The thickness of the deposited ethanethiol ice reduces in each of the repeated cycles. The figure is blown up and presented in two sets of wavenumbers i.e. CH str and SH str region (top), fingerprint region (bottom), so that the spectral features are clearly visible.

**Figure 4:** Ethanethiol ice formed at 4 K and (a) subsequently warmed to 120 K (crystalline), (b) warmed to 126 K (amorphous) and (c) cooled to 118 K (crystalline).

**Table 1:** Observed peak positions in the IR spectra of ethanethiol ices in both amorphous and crystalline phases at three different temperatures. Note: certain assignments are made using[16].

| Peak position (cm$^{-1}$) | | | Assignment |
|---|---|---|---|
| Amorphous (4 K) | Crystalline (120 K) | Amorphous (125 K) | |
| 655 | 656, 654 | 655 | CS stretching |
| 733 | 741, 730 | 734 | CH$_2$ rocking |
| 782 | 780 | 780 | CH$_2$ rocking |
| 864 | 865, 874 | 866 | SH out of plane bending |
| 969 | 967 | 969 | CC stretching |
| 1030 | 1031 | 1030 | CH$_2$ rocking |
| 1050 | 1048 | 1050 | CH$_2$ rocking |
| 1094 | 1099, 1092 | 1095 | CH$_3$ bending |
| 1250 | 1251 | 1250 | CH$_2$ twisting |
| 1271 | 1272 | 1271 | CH$_2$ wagging |
| 1430 | 1425, 1428 | 1430 | CH$_2$ scissoring |
| 1448 | 1446 | 1448 | CH$_3$ bending |
| 2528 | 2526, 2541 | 2538 | SH stretching |
| 2864 | 2,86,62,841 | 2865 | CH$_3$ symmetric stretching |
| 2924 | 2926 | 2923 | CH$_2$ symmetric stretching |
| 2959 | 2962 | 2960 | CH$_3$ asymmetric stretching |
| 2975 | 2980, 2974 | 2976 | CH$_2$ asymmetric stretching |

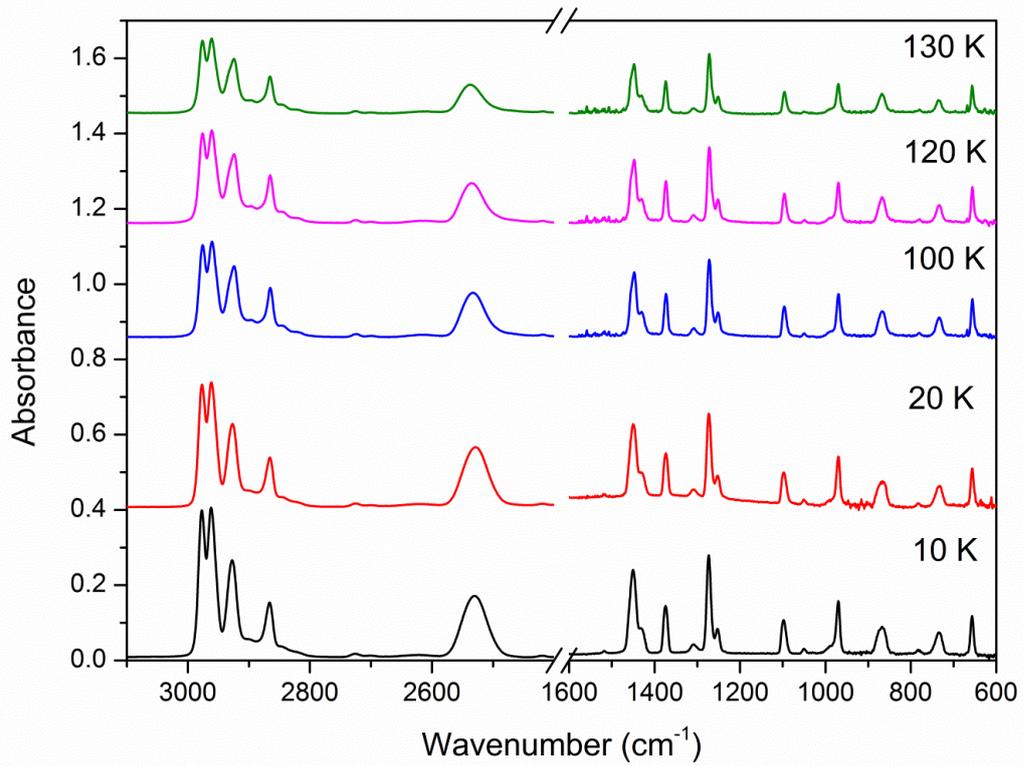

**Fig. 1a**

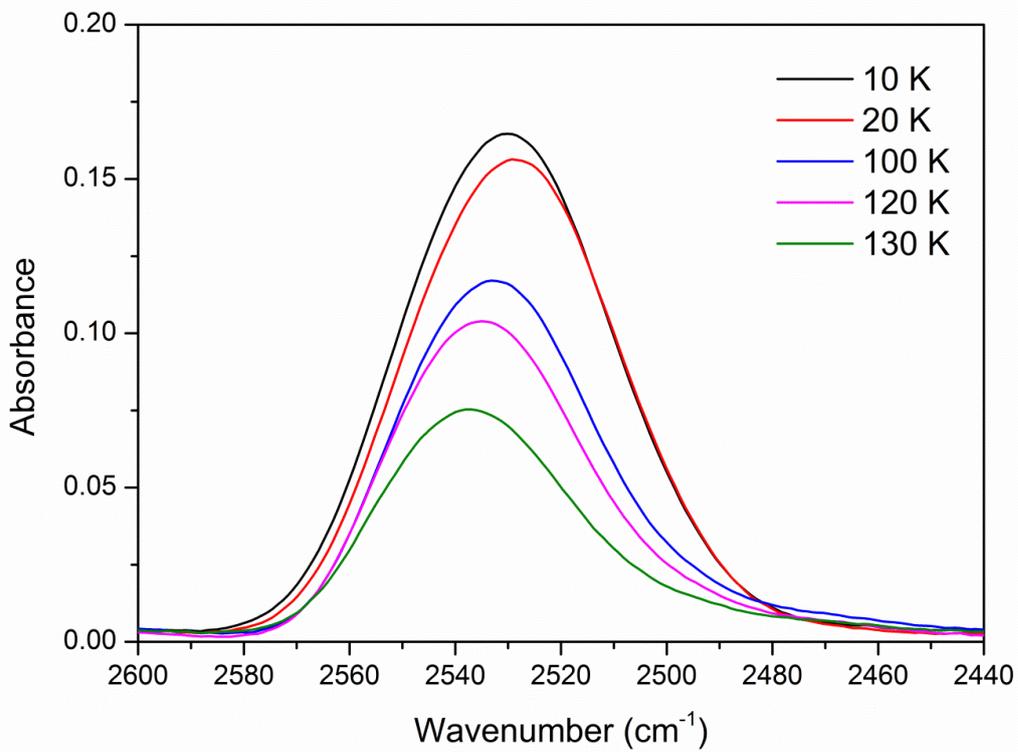

**Fig. 1b**

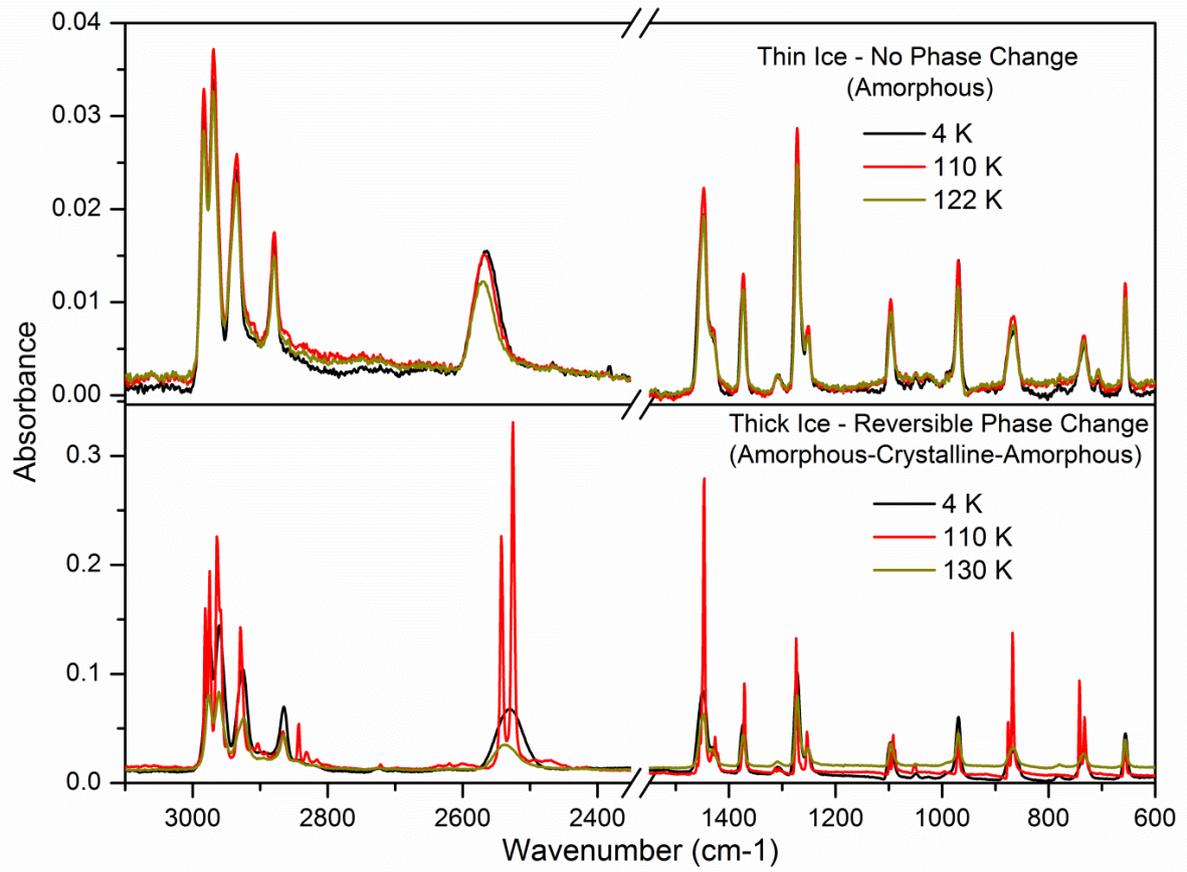

**Fig. 2a**

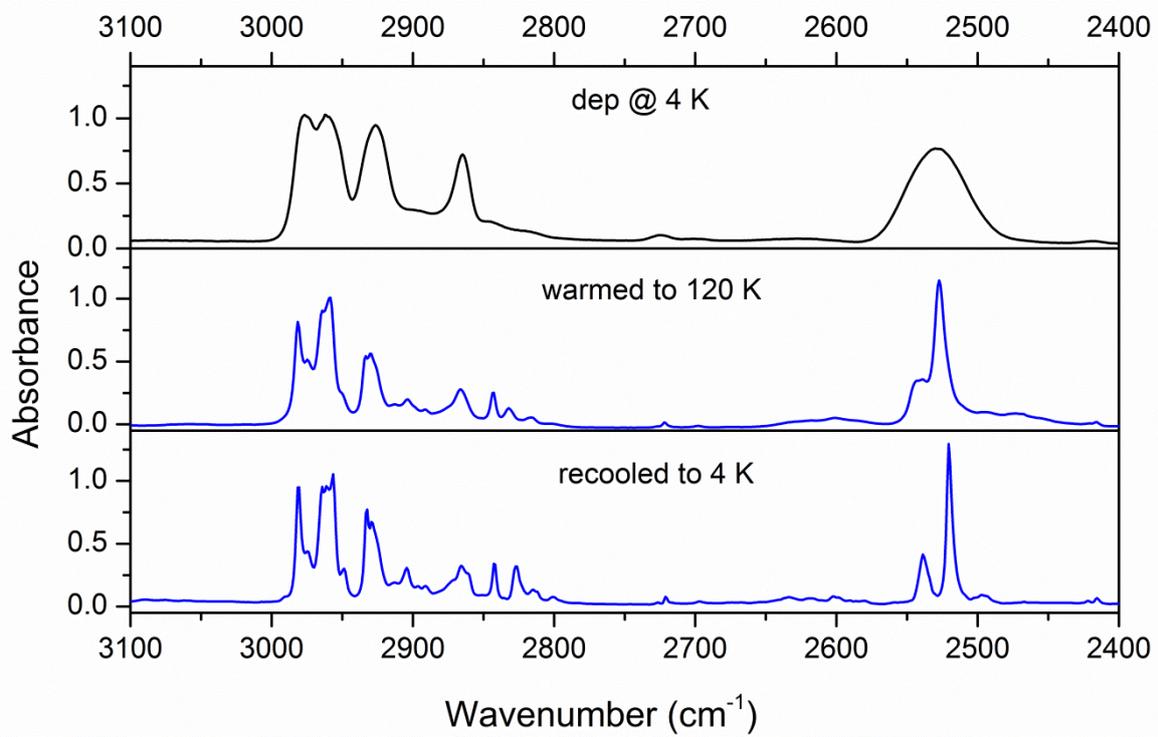
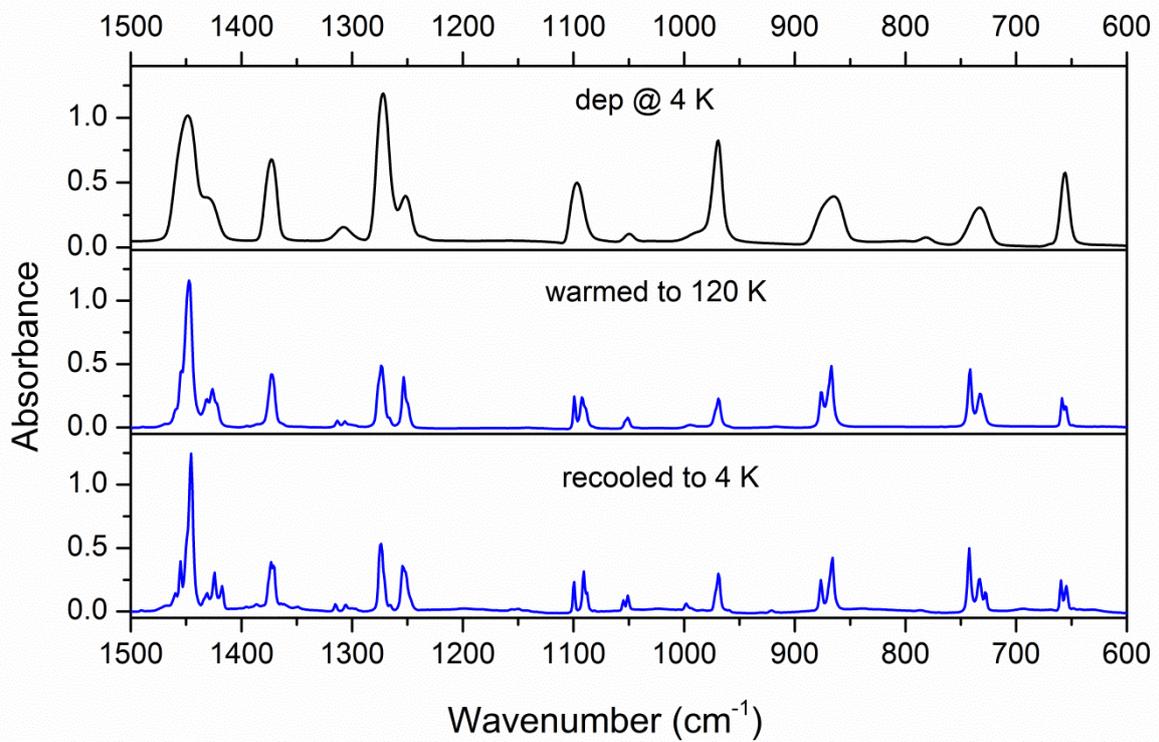

**Fig. 2b**

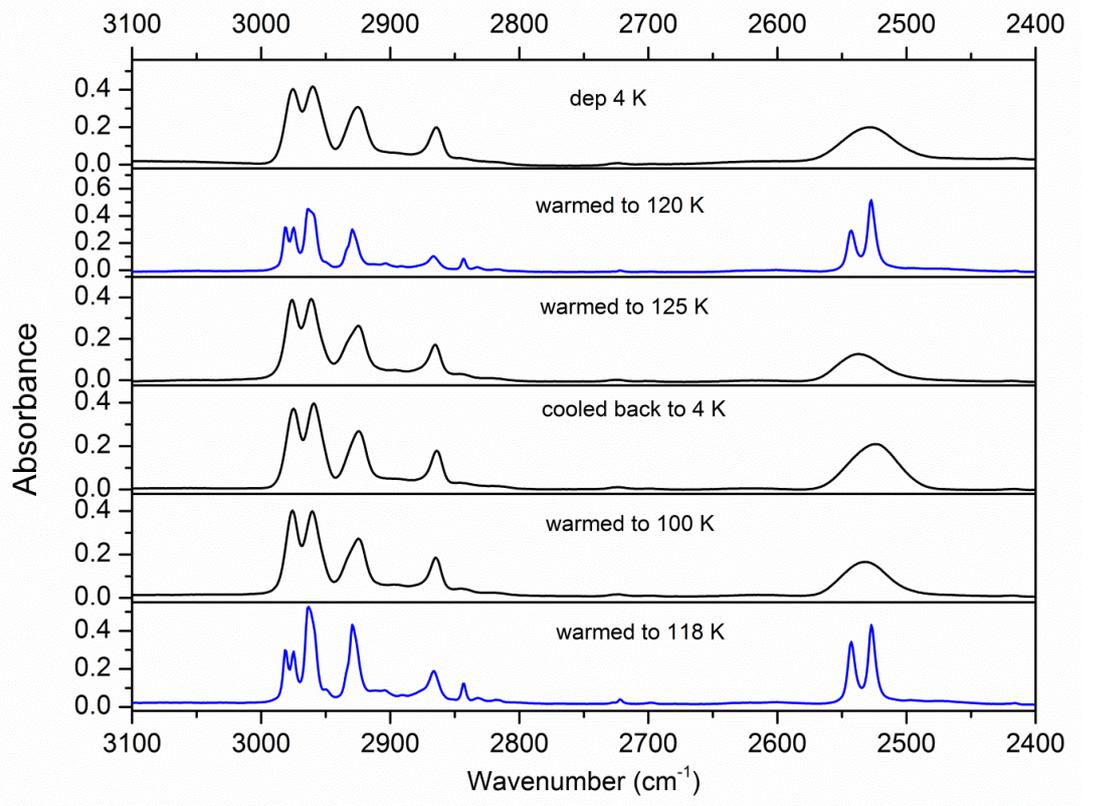
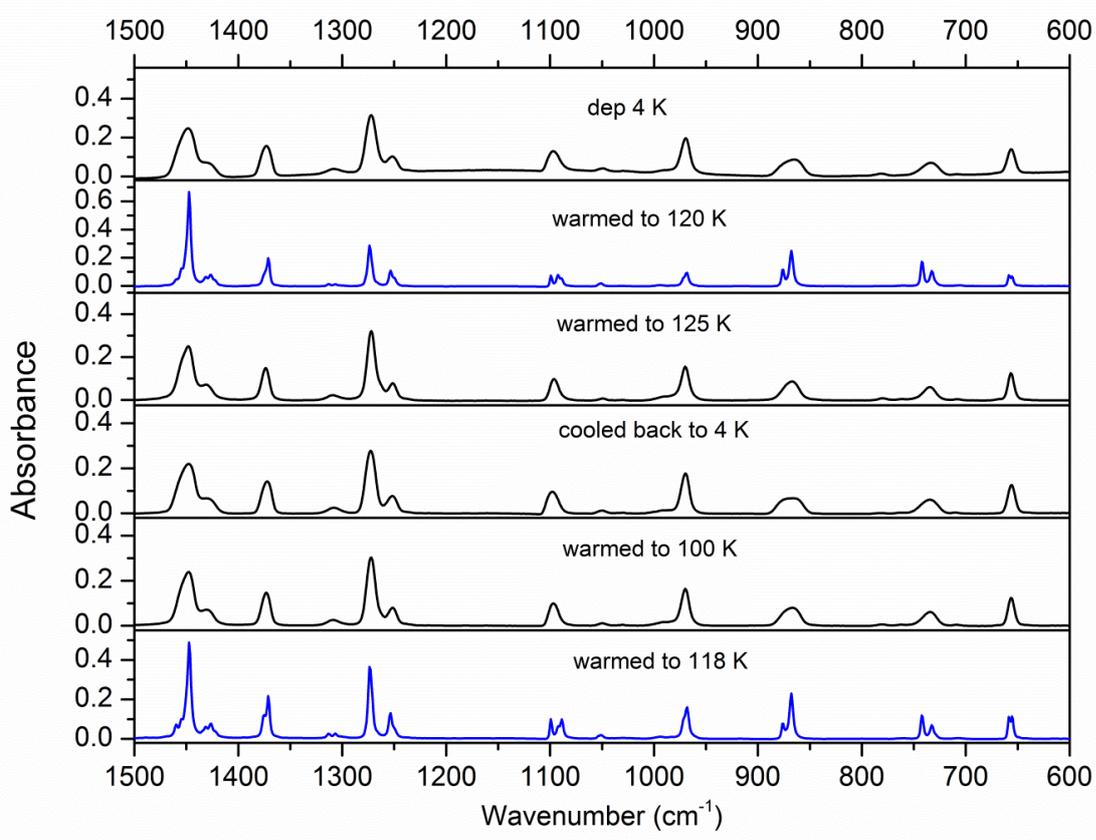

**Fig. 3a**

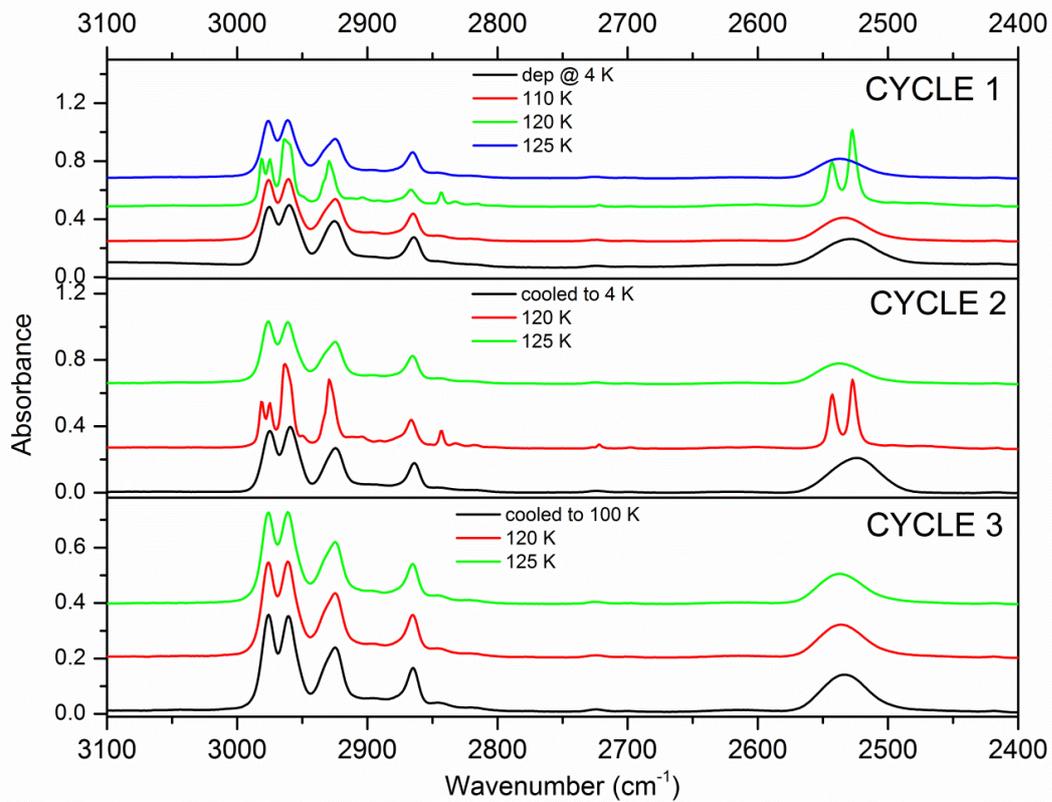

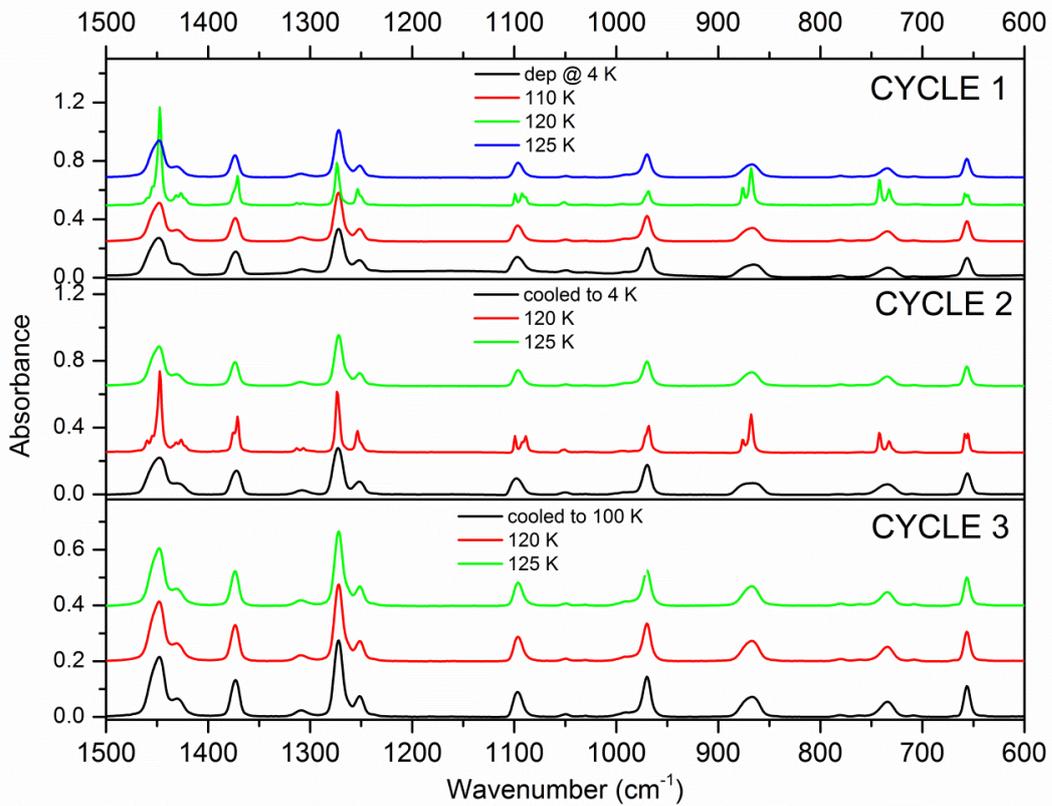

**Fig. 3b**

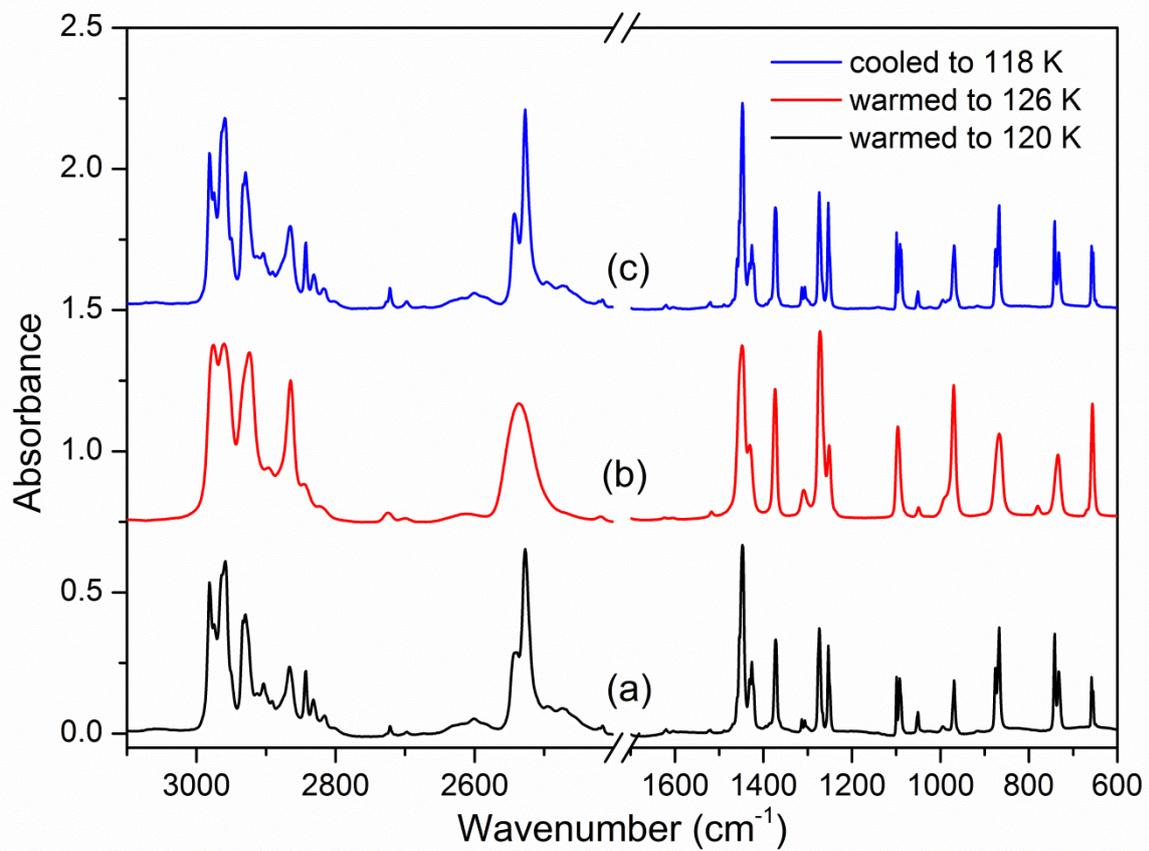

**Fig. 4**